# Role of geometrical cues in neuronal growth

Joao Marcos Vensi Basso[1], Ilya Yurchenko[1], Marc Simon[1], Daniel J. Rizzo[1,2], Cristian Staii[1,*]

1. Department of Physics and Astronomy, Center for Nanoscopic Physics, Tufts University, Medford, Massachusetts 02155, *USA*
2. Current address: Department of Physics, University of California, Berkeley, California, 94720, *USA*

 [*] Corresponding Author E-mail: Cristian.Staii@tufts.edu

## Abstract

Geometrical cues play an essential role in neuronal growth. Here, we quantify axonal growth on surfaces with controlled geometries and report a general stochastic approach that quantitatively describes the motion of growth cones. We show that axons display a strong directional alignment on micro-patterned surfaces when the periodicity of the patterns matches the dimension of the growth cone. The growth cone dynamics on surfaces with uniform geometry is described by a linear Langevin equation with both deterministic and stochastic contributions. In contrast, axonal growth on surfaces with periodic patterns is characterized by a system of two generalized Langevin equations with both linear and quadratic velocity dependence and stochastic noise. We combine experimental data with theoretical analysis to measure the key parameters of the growth cone motion: angular distributions, correlation functions, diffusion coefficients, characteristics speeds and damping coefficients. We demonstrate that axonal dynamics displays a cross-over from an Ornstein-Uhlenbeck process to a non-linear stochastic regime when the geometrical periodicity of the pattern approaches the linear dimension of the growth cone. Growth alignment is determined by surface geometry, which is fully quantified by the deterministic part of the Langevin equation. These results provide new insight into the role of curvature sensing proteins and their interactions with geometrical cues.

# 1. INTRODUCTION

Neuronal cells are the primary working units of the nervous system. A single neuron is an extremely specialized cell that develops two types of processes: axons and dendrites [Fig. 1(a)]. These processes grow and make connections with other neurons thus wiring up the nervous system. Once the neural circuits are formed, electrical signals are transmitted among neurons through functional connections (synapses) made between axons and dendrites. During the development of the brain axons actively navigate over distances of the order of 10-100 cell diameters in length to find target dendrites from other neurons and to form neural circuits [1, 2]. Axonal motion is guided by the growth cone, a dynamic sensing unit located at the leading edge of the axon. The growth cones consistently follow specific pathways through a complex and changing environment by responding to multiple guidance cues [1-4].

In recent years many intercellular signaling processes that control growth cone migration have been investigated in great detail [2, 5-7], and there is now a considerable amount of information about the molecular machinery that regulates these processes. For example, there are several comprehensive models that describe receptor-ligand interactions, as well as changes in the cellular cytoskeleton in response to biochemical cues from the environment or from other neurons [1-6, 8]. However, much less is known about how growth cones sense, react and move in response to *mechanical* or *geometrical* cues.

Axonal alignment and directional growth have been demonstrated in many *in vitro* studies of neuronal growth on micro-patterned substrates [3, 5, 9-13]. An important research direction in this area is to precisely measure and characterize axonal dynamics on surfaces with periodic micro-patterns. Previous experiments have demonstrated that neurons cultured on surfaces with periodic geometrical patterns display a significant increase in the total length of axons, as well as axonal alignment along preferred growth directions [10, 12-14]. Despite these advances there remain many important questions about the actual mechanisms that determine axonal alignment in response to surface geometry, such as: (i) how the growth cones sense geometrical features; (ii) how adhesion forces are generated as growth cones advance, retract or turn; (iii) what biophysical processes lead to the symmetry - breaking events that determine axonal directionality and alignment. Furthermore, many of the previous studies provide mainly qualitative or semi-quantitative descriptions of axonal migration and alignment. A fully quantitative picture of neuronal growth on surfaces with controlled geometries is still missing.

Several types of stochastic processes are involved in the growth cone dynamics including, neuron-neuron signaling, fluctuating weak environmental biochemical cues, biochemical reactions taking place in the growth cone, and polymerization/de-polymerization processes involved in actin and microtubule dynamics [1, 2, 6, 15]. Therefore, the motion of each *individual* growth cone cannot be predicted. However, the behavior of ensembles of growth cones belonging to axons from many neurons can be described quantitatively by Langevin-type stochastic differential equations [12, 16, 17]. Quite generally the dynamics of the growth cone on micro-patterned surfaces is controlled both by a deterministic component (tendency to grow in certain preferred directions imparted by surface geometry), and a random deviation from these growth directions due to stochastic processes.

Langevin and Fokker-Planck equations provide a powerful framework for modeling the interplay between the deterministic and stochastic components of biased random motion. When applied to axonal growth, these equations enable accurate prediction of growth cone dynamics, and provide a systematic approach for analyzing the respective roles played by external biochemical, mechanical, and geometrical cues. In our previous work we have used a theoretical model based on the Fokker-Planck (FP) equation to quantify axonal growth on glass [17] and on surfaces with engineered, ratchet-like topography (asymmetric tilted nanorod, or nano-ppx surfaces) [12]. We have demonstrated axonal alignment towards a single dominant direction on nano-ppx surfaces and have measured the diffusion coefficient on these surfaces. We have shown that the angular distributions originate from the axon-surface interaction forces that ultimately produce a "deterministic torque", which aligns the growth cones along certain preferred directions of growth [12].

A general characteristic of the stochastic models is that they can be used to obtain key dynamical parameters that characterize the cellular motion such as: diffusion (cell motility) coefficients, mean square displacements, velocity and angular correlation functions [16-22]. Moreover, these phenomenological models can serve as a basis for quantifying cell-cell and cell-surface interactions, which can then be explained in terms of biophysical and biochemical processes taking place inside the cell. Ultimately, this approach could lead to a detailed understanding of the basic principles that govern the formation of neural circuits, which is one of the long-term goals in neuroscience.

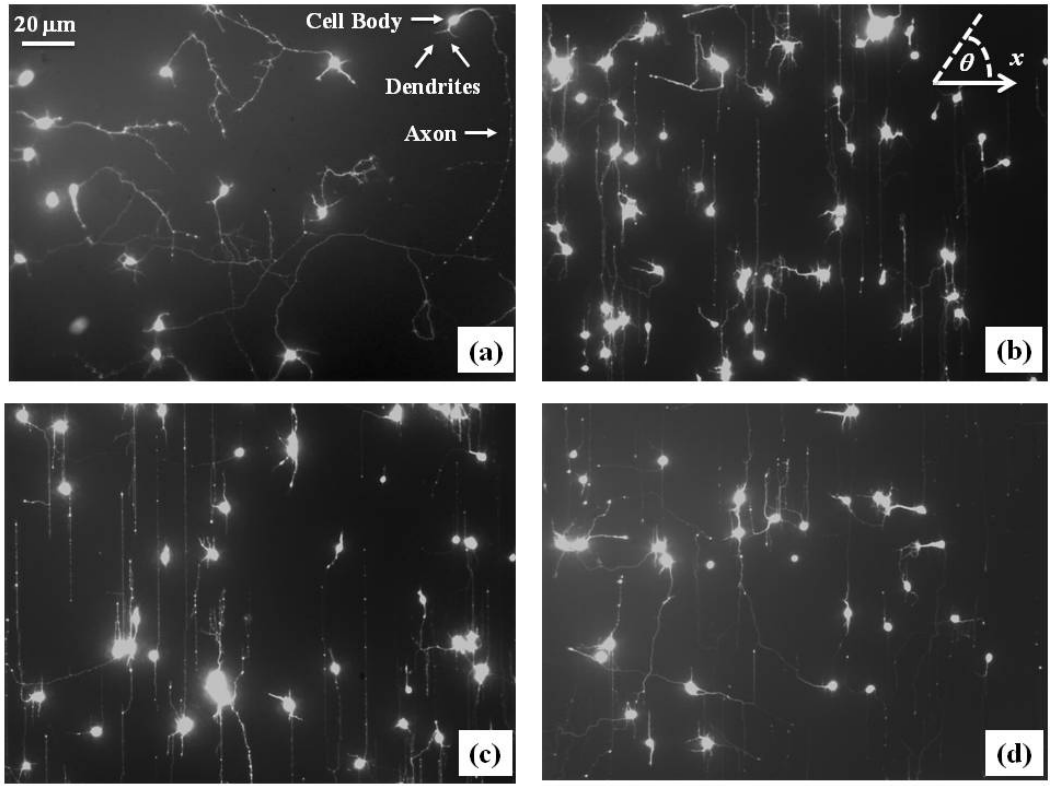

**FIG. 1.** Examples of cultured cortical neurons on PDL coated glass (a) (flat surface without micro-patterns), and PDL coated PDMS surfaces (b-d) with periodic micro-patterns shown in Fig. 2. The main structural components of a neuronal cell are labeled in Fig. 1(a). The angular coordinate $\theta$ used in this paper is defined in the inset of Fig. 1(b). All angles are measured with respect to the $x$ axis, defined as the axis perpendicular to the direction of the PDMS patterns (see Fig. 2). The pattern spatial period is: $d = 5$ μm in Fig. 1(b); $d = 3$ μm in Fig. 1(c); $d = 1.5$ μm in Fig. 1(d). Images (a-d) are captured 48 hrs after neuron plating.

In this paper we present a systematic experimental and theoretical investigation of axonal growth for cortical neurons cultured on poly-D-lysine (PDL) coated glass, and several types of PDL coated polydimethylsiloxane (PDMS) surfaces with periodic micro-patterns (Fig. 1). We create PDMS surfaces with periodic features (parallel ridges separated by troughs), each surface being characterized by a different value of the geometrical parameter *d* (Fig.2) defined as the distance between two neighboring ridges (and referred to as the *pattern spatial period* throughout the text). We demonstrate that axons tend to grow parallel to the surface micro-patterns, and that the highest degree of axonal alignment occurs when the pattern spatial period approaches the linear dimension of the growth cone (Fig. 1 and Fig. 2).

We show that experimental data for neurons grown on PDL coated glass is well-described by a linear Langevin equation with white noise, i.e. an Ornstein-Uhlenbeck (OU) process. Furthermore, growth on micro-patterned PDMS surfaces is also described by an OU process *only* for the cases in which the pattern spatial period *d* is much larger (or much smaller) than the dimension of the growth cone. On the other hand, neuronal growth on PDMS surfaces where the pattern spatial period matches the dimension of the growth cone cannot be described by an OU process. We demonstrate that the growth dynamics on these surfaces is described by non-linear Langevin equations containing linear and quadratic velocity terms, angular orientation terms, and stochastic terms with Gaussian white noise. This model fully accounts for the experimental data, including growth speeds, axonal alignment, velocity correlation functions, and angular distributions. Using this model, we extract from the experimental data the main parameters that characterize the motion of the growth cone on PDMS patterned surfaces: velocity damping coefficients, deterministic accelerations and terminal velocities. We show that these parameters are controlled solely by the surface geometry (pattern spatial period) and represent a measure of the cell-cell and cell-surface interactions. We therefore demonstrate that the *deterministic part* of the Langevin equation quantifies the effects of *surface geometry*, which is ultimately responsible for directional growth (alignment with the pattern). These results are important for the fundamental understanding of how geometrical cues influence axonal growth and dynamics, as well as for bioengineering novel substrates to control neuronal growth.

This paper is structured as follows. In Section II and Section III we present respectively, details of the experimental procedure and data analysis. Axonal dynamics on glass is treated in Section IV. In Section V we present the experimental data and the theoretical model for growth on patterned PDMS surfaces. Section VI contains a detailed discussion of the experimental results, predictions of the theoretical model, and comparisons with previous results of cell motility reported for neurons and other types of cells. We present the conclusions in Section VII.

## II. EXPERIMENTAL DETAILS

All cells used in this work are cortical neurons obtained from embryonic day 18 rats. For cell dissociation and culture we have used established protocols detailed in our previous work [9, 12, 17, 23-25]. Cortical neurons were cultured either on poly-D-lysine (PDL) coated glass or on micro-patterned polydimethylsiloxane (PDMS) substrates coated with PDL.

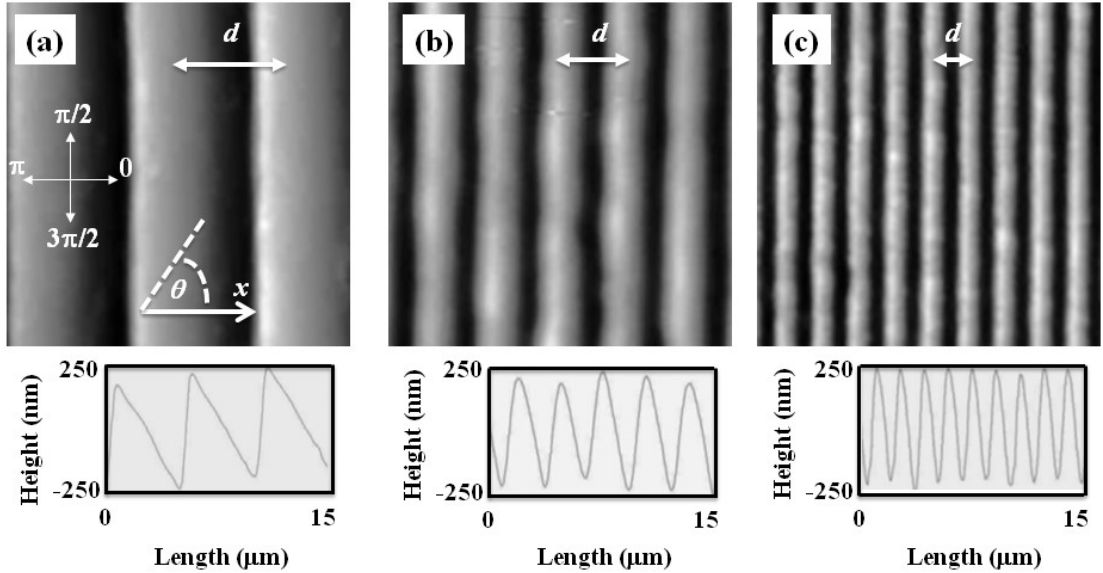

**FIG. 2.** Topographic Atomic Force Microscope (AFM) images of PDL coated PDMS patterns corresponding to the neuron growth surfaces shown in Fig. 1(b-d). Fig. 2(a) defines the angular coordinate $\theta$ used in this paper. The *x* axis is defined as the axis perpendicular to the direction of the PDMS patterns. The directions corresponding to $\theta = 0, \pi/2, \pi,$ and $3\pi/2$ are also shown in Fig. 2(a). The pattern spatial period *d* defined as the distance between two neighboring ridges is shown for each pattern. Examples of AFM line scans obtained across each surface are shown at the bottom of each figure. The line scans demonstrate that the patterns are periodic in the *x* direction, and have a constant depth of approximately 0.5 μm. The values for *d* are: *d =5* μm in Fig. 2(a); *d =3* μm in Fig. 2(b); *d =1.5* μm in Fig. 2(c).

The micro-patterns on PDMS surfaces consist of periodic features (parallel ridges separated by troughs). Each surface is characterized by a different value of the pattern spatial period *d,* defined as the distance between two neighboring ridges (Fig.2). To obtain these periodic patterns we used a simple fabrication method based on imprinting diffraction grids with different grating constants onto PDMS substrates [25]. The direction of the patterns is shown in Fig. 2 by the parallel white stripes (ridges), as well as by the parallel black stripes (troughs).

The surfaces were then spin-coated with PDL (Sigma-Aldrich, St. Louis, MO) solution of concentration 0.1 mg/mL. Neuronal cells were imaged using an MFP3D atomic force microscope (AFM) equipped with a BioHeater closed fluid cell, and an inverted Nikon Eclipse Ti optical microscope (Micro Video Instruments, Avon, MA). Fluorescence images were acquired using a standard Fluorescein isothiocyanate -FITC filter: excitation: 495 nm and emission: 521 nm. To acquire the fluorescence images the neurons were incubated for 30 minutes at 37º C with 50 nM Tubulin Tracker Green (Oregon Green 488 Taxol, bis-Acetate, Life Technologies, Grand Island, NY) in PBS. The samples were then rinsed with PBS and re-immersed in PBS solution for imaging [12, 23-25].

## III. DATA ANALYSIS

Growth cone position, axonal length, and angular distributions have been measured and quantified using ImageJ (National Institute of Health). The displacement of the growth cone was obtained by measuring the change in the center of the growth cone position. To measure the growth cone velocities the samples were imaged every $\Delta t = 5$ min for a total period between 30 min - 2hrs. The 5 min time interval between measurements was chosen such that the typical displacement $\Delta \vec{L}$ of the growth cone in this interval satisfies two requirements: a) is larger than the experimental precision of our measurement (~ 0.1 μm) [17]; b) the ratio $\Delta \vec{L}/\Delta t$ accurately approximates the instantaneous velocity $\vec{V}$ of the growth cone.

The instantaneous velocity $\vec{V}(t)$ for each growth cone at time $t$ is determined by using the formula:

$$\vec{V}(t) = \frac{\Delta \vec{L}}{\Delta t} = \frac{\vec{r}(t+\Delta t) - \vec{r}(t)}{\Delta t}. \qquad (1)$$

where $\vec{r}(t)$ is the position vector of the growth cone at time $t$, and $\Delta \vec{L}$ is the net displacement of the same growth cone during the time interval $\Delta t = 5$ min between the measurements. The speed is defined as the magnitude of the velocity vector: $V(t) = |\vec{V}(t)|$, and the growth angle $\theta(t)$ is measured with respect to the $x$ axis (growth angle and the $x$ axis are defined in Fig. 2(a)).

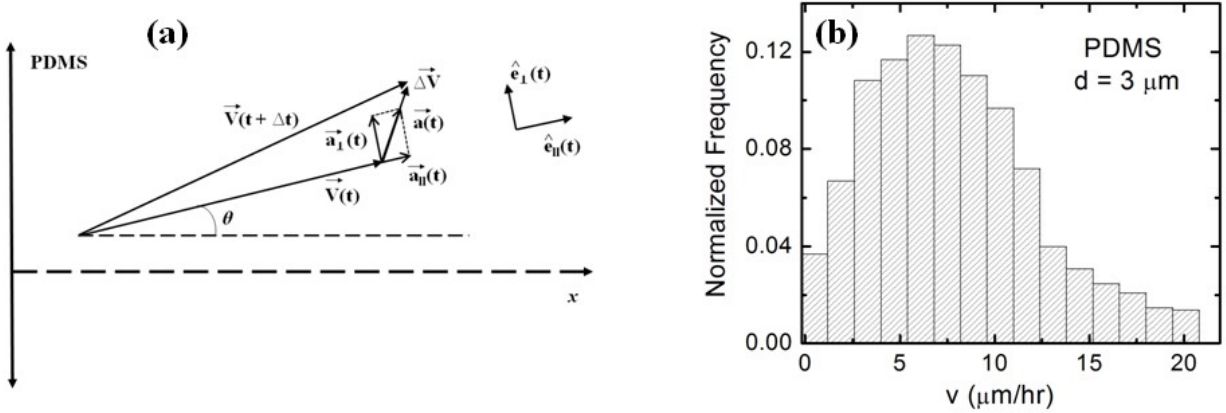

**FIG. 3.** (a) Schematic of the coordinate system. (b) Example of normalized speed distribution for growth cones measured on PDMS substrates with spatial period $d = 3$ μm. The histogram was obtain by plotting the speed measured for N = 117 different growth cones.

The acceleration for each growth cone is calculated at each point of its trajectory by finite differences of the cell velocities, and measured in a moving coordinate frame attached to the growth cone (Fig. 3(a)). The two unit vectors $\hat{e}_\parallel(t)$ and $\hat{e}_\perp(t)$ of the moving frame are oriented parallel (and respectively perpendicular) to the instantaneous velocity $\vec{V}(t)$ of the growth cone (Fig. 3(a)). To measure the deterministic and stochastic parts of the acceleration we have used the method of conditional averaging of experimental data as described in ref. [18, 26]. For each image the range of growth cone speeds was divided into 15 intervals of equal size $|\Delta \vec{V}_0|$ (Fig. (3b)). Experimental data (Fig. 1) shows that over a distance of ~ 20 μm the axons can be approximated by straight line segments, with a high degree of accuracy. Therefore, to obtain the angular distributions (Fig. 4 and Fig. S1 in Supplemental Material [25]) we have tracked all axons using ImageJ and then partitioned them into segments of 20 μm in length. Next, we have recorded the angle that each segment makes with the $x$ axis (Fig. 2). The total range $[0, 2\pi]$ of growth angles was divided into 18 intervals of equal size $\Delta \theta_0 = \pi/9$ (Fig 4).

To obtain the deterministic parts of the growth cone acceleration we have averaged the parallel and perpendicular components of the acceleration (i.e. the components along the $\hat{e}_{\parallel}(t)$ and $\hat{e}_{\perp}(t)$ directions in Fig. 3(a)) within each speed $|\Delta \vec{V}_0|$ (or angle $\Delta \theta_0$) interval, according to the equations [18, 26]:

$$a_{d,\parallel}(\vec{V},t) = \left\langle \left(\frac{\Delta \vec{V}}{\Delta t}\right) \cdot \hat{e}_{\parallel}(t) \right\rangle_C \qquad (2)$$

$$a_{d,\perp}(\vec{V},t) = \left\langle \left(\frac{\Delta \vec{V}}{\Delta t}\right) \cdot \hat{e}_{\perp}(t) \right\rangle_C \qquad (3)$$

Here $a_{d,\parallel}$ and $a_{d,\perp}$ are respectively, the parallel and perpendicular components of the deterministic acceleration $a_d$. The subscript $C$ stands for the conditional average:

$$C: \begin{cases} |\vec{V}_i(t) - \vec{V}(t)| \leq |\Delta \vec{V}_0| \\ |\theta_i(t) - \theta(t)| \leq \Delta \theta_0 \end{cases} \qquad (4)$$

where $\vec{V}_i$ and $\theta_i$ are, respectively, the velocity and the angle of the i$^{th}$ growth cone.

The stochastic terms in the parallel and perpendicular directions are obtained by calculating the variance of the distribution for each component of the acceleration within each speed and angle interval, according to the formulas [18, 26]:

$$\Gamma^2{}_{\parallel} = \left\langle \left(a_{\parallel}(\vec{V},t) - \langle a_{\parallel}(\vec{V},t)\rangle\right)^2 \right\rangle_C = \left\langle \left(a_{\parallel}(\vec{V},t) - a_{d,\parallel}(\vec{V},t)\right)^2 \right\rangle_C \qquad (5)$$

$$\Gamma^2{}_{\perp} = \left\langle \left(a_{\perp}(\vec{V},t) - \langle a_{\perp}(\vec{V},t)\rangle\right)^2 \right\rangle_C = \left\langle \left(a_{\perp}(\vec{V},t) - a_{d,\perp}(\vec{V},t)\right)^2 \right\rangle_C \qquad (6)$$

We found that the cross-correlation terms for each component of the acceleration are much smaller (by at least one order of magnitude) than the stochastic terms given by Eqn. (5) and (6) (see Supplemental Material [25]). We therefore neglect the mixed stochastic terms in our analysis and include only the stochastic contributions given by Eq. (5) and Eq. (6). Experimental data also shows that the stochastic contributions for both the parallel and perpendicular components of the acceleration fluctuate around their average values with zero correlations (see Fig. S4 in Supplemental Material [25]). Consequently we model the stochastic terms as being represented by uncorrelated Gaussian white noise sources with zero mean [12, 16, 17, 19, 20].

Experimentally, the velocity autocorrelation function is obtained according to the formula [19, 20]:

$$C_V(t) = \frac{1}{N} \cdot \sum_{i=1}^{N} \left(\vec{V}_i(t) \cdot \vec{V}_i(0)\right) \qquad (7)$$

where $N$ is the total number of growth cones and $\vec{V}_i(t)$ represents the velocity of the i$^{th}$ growth cone at time $t$.

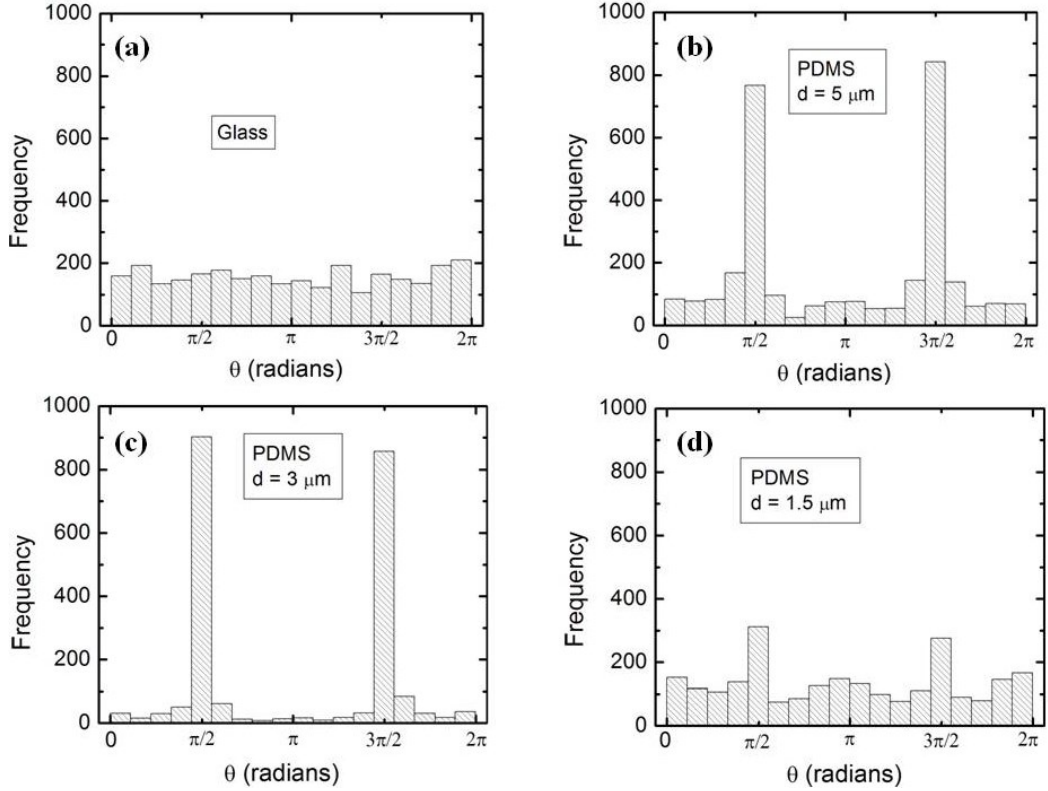

**FIG. 4.** Experimental angular distributions for axonal growth on different types of surfaces (shown in Fig. 1 and Fig. 2). The vertical axis (labeled Frequency) represents the number of axonal segments, each one of 20 μm in length (see DATA ANALYSIS section). (a) Data for PDL coated glass (N = 2856 axon segments); (b) Data for PDL coated PDMS substrates with spatial period $d = 5$ μm (N = 2940 axon segments); (c) Data for PDL coated PDMS substrates with spatial period $d = 3$ μm (N = 2242 axon segments); (d) Data for PDL coated PDMS substrates with spatial period $d = 1.5$ μm (N = 2545 axon segments). The axons of neurons cultured on PDMS surfaces with spatial period $d = 5$ μm (Fig. 4(b)) and $d = 3$ μm (Fig. 4(c)) display strong directional alignment with the surface patterns (peaks at $\theta = \pi/2$ and $\theta = 3\pi/2$, see Fig. 1 and Fig. 2). The axons of neurons cultured on PDMS surfaces with spatial period $d = 1.5$ μm show a lower degree of alignment (Fig. 4(d)), while the axons of neurons cultured on glass show no directionality (Fig. 4(a)).

## IV. NEURONAL GROWTH ON GLASS SURFACES

In previous work [17] we have analyzed axonal dynamics on PDL coated glass substrates using the Fokker-Planck equation. Here we show that the growth on these substrates is described by an Ornstein-Uhlenbeck (OU) process, and measure the dynamical parameters that describe this process. Fig. 1(a) shows an example of axonal growth on PDL coated glass, obtained 48 hrs after neuronal plating. The angular distribution for growth on glass is shown in Fig. 4(a).

The OU process is described by the following linear Langevin equation for the velocity $\vec{V}$ [19, 20, 27]:

$$\frac{d\vec{V}}{dt} = -\gamma_g \cdot \vec{V} + \vec{\Gamma}(t) \qquad (8)$$

The first term in Eq. (8) represents the deterministic term, and $\gamma_g$ is a constant damping coefficient. The second term $\vec{\Gamma}(t)$ represents the stochastic change in velocity, which is described by Gaussian white noise [17, 19, 20, 27]. In the absence of the stochastic term the velocity would decay exponentially with a characteristic time: $\tau_g = 1/\gamma_g$ (throughout this paper we will use the subscript "g" to denote the parameters for glass). Since the axonal motion takes place in two spatial dimensions, equation (8) implies the following expressions for the axonal mean square length $<\vec{L}^2(t)>$, and the growth cone velocity autocorrelation function $C_V(t)$ as functions of time [19, 20, 22]:

$$\langle \vec{L}^2(t) \rangle = 4D \cdot t - \frac{4D}{\gamma_g} \cdot (1 - e^{-\gamma_g t}) \qquad (9)$$

$$C_V(t) \equiv \langle \vec{V}(t) \cdot \vec{V}(0) \rangle = 2D \cdot \gamma_g \cdot e^{-\gamma_g t} \qquad (10)$$

where $D$ is the cell random motility coefficient [1, 16, 17, 20-22]. At the level of cell populations the random motility coefficient is analogous to diffusion coefficient of the OU process [20, 22]. In this paper we will refer to $D$ as diffusion coefficient, as it is customary in literature [1, 16, 17, 20-22].

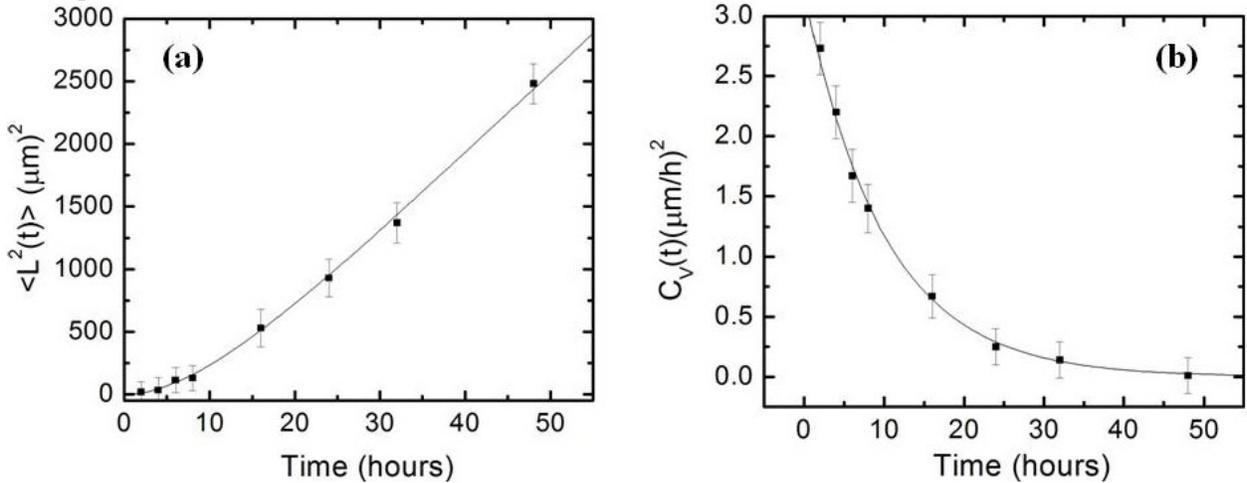

**FIG. 5.** (a) Data points: experimentally measured average axonal length square vs. time. The continuous curve represents the fit of the data points with the prediction of the theoretical model based on the Ornstein-Uhlenbeck process (Eq. (9)). (b) Experimentally measured velocity correlation function (data points) as a function of time. The continuous curve represents the fit to the data with Eq. (10) (prediction of the theoretical model based on the Ornstein-Uhlenbeck process). Each data point in Fig. 5(a) and Fig. 5(b) was obtained by measuring between N = 120 and N = 170 different axons (corresponding to 5-10 different fluorescent images per time data point). Error bars in both figures indicate the standard error of the mean. The fit of the data with Eq. (9) and Eq. (10) give the diffusion coefficient $D$ and the constant damping coefficient $\gamma_g$ for the Ornstein-Uhlenbeck process (see text).

To find the two parameters: $D$ and $\gamma_g$, which characterize the OU process we performed a time series measurement of axonal lengths and velocities (for $t$ = 2, 4, 6, 8, 16, 24, 32, 48 hrs) for neurons cultured on glass. From this data we have experimentally measured the average axonal length square $<\vec{L}^2(t)>$ and velocity autocorrelation function $C_V(t)$ (Eq. (7)) at each time $t$ (see DATA ANALYSIS). Fig. 5(a) shows the experimental data for the mean square displacement vs.

time, together with the fit of the data (continuous curve) with Eq. (9), which represents the theoretical prediction based on the OU process. Fig. 5(b) shows the experimental data for the velocity autocorrelation function vs. time, together with the fit of the data (continuous curve) obtained by using Eq. (10) (theoretical prediction based on the OU process). From the fit of the data in Fig. 5(a) and Fig. 5(b) we obtain the following values for the diffusion coefficient $D$ and for the constant damping coefficient $\gamma_g$ : $D = (16 \pm 2)\,\mu m^2/hr$ and $\gamma_g = (0.1 \pm 0.05)\,hr^{-1}$. The value for the diffusion coefficient agrees with our previous results for neurons on glass [17]. From the value of the damping coefficient we find a characteristic time for the exponential decay of the velocity autocorrelation function: $\tau_g = 1/\gamma_g \approx 10\,hr$.

Since the axonal growth on glass is described by an OU process (Eqs. 8-10), we can relate $D$ and $\gamma_g$ with the mean square velocity for neuronal growth on glass via [19, 20, 22]:

$$D = \frac{\langle V_c^2 \rangle \cdot \tau_g}{2} = \frac{\langle V_c^2 \rangle}{2\gamma_g} \tag{11}$$

Using the experimentally measured values for $D$ and $\gamma_g$, Eqn. (11) predicts the following value for the characteristic speed of neuronal growth on glass:

$$V_g \equiv \sqrt{\langle V_c^2 \rangle} = \sqrt{2D\gamma_g} \approx 1.8\,\mu m/hr \tag{12}$$

In conclusion, we found that neuronal growth on PDL coated glass is well-described by an Ornstein-Uhlenbeck process (linear Langevin equation with Gaussian white noise). There is no preferred directionality for the axonal growth (Fig. 1(a) and Fig 4(a)). By fitting the experimental data with the theoretical OU model (Fig. 5) we measure the fundamental dynamical parameters for neuronal growth on glass: diffusion coefficient $D$, constant damping coefficient $\gamma_g$, characteristic time $\tau_g$, and use these values to calculate a characteristic speed of axonal growth $V_g$. These values are in agreement with our previous results obtained by using the Fokker-Planck equation for describing neuronal growth on glass surfaces [17].

## V. NEURONAL GROWTH ON MICRO-PATTERNED PDMS SURFACES

We cultured neurons on PDL coated PDMS surfaces with controlled geometrical patterns (i.e. periodic parallel ridges separated by troughs). The surfaces differ by the value of the pattern spatial period $d$ (distance between two neighboring ridges, Fig. 2). We analyze growth for surfaces with spatial periods: $d$ = 10 μm, 5 μm, 3 μm, 1.5 μm and 0.5 μm. Images of axonal growth on these surfaces and the corresponding angular distributions are shown, respectively, in Fig. 1(b) and Fig. 4(b) (for $d$ = 5 μm), Fig. 1(c) and Fig. 4(c) (for $d$ = 3 μm), and Fig. 1(d), and Fig. 4(d) (for $d$ = 1.5 μm). Examples of AFM topography images obtained on these surfaces are shown in Fig. 2. Images of axonal growth on surfaces with $d$ = 10 μm and $d$ = 0.5 μm, as well as the corresponding angular distributions are shown in Fig. S1 in the Supplemental Material [25].

The experimental data shows that there is no preferred directionality for axonal growth on PDMS surfaces with pattern spatial periods $d$ = 10 μm and $d$ = 0.5 μm (Fig. S1 in the Supplemental Material [25]). In contrast, neurons cultured on surfaces with $d$ = 5 μm and $d$ = 3 μm display strong directional alignment with the surface patterns (Fig. 1(b-c) and Fig. 4(b-c)). Axons grown on surfaces with $d$ = 1.5 μm display a lower degree of alignment with the surface pattern (Fig. 1(d) and Fig. 4(d)).

**Neuronal growth on micro-patterned PDMS surfaces with spatial periods $d$ = 5 μm, $d$ = 3 μm, and $d$ = 1.5 μm.** We focus first on analyzing neuronal growth on PDMS surfaces with the pattern spatial periods $d$ = 5 μm, $d$ = 3 μm, and $d$ = 1.5 μm for which experimental data shows axonal alignment along the surface patterns (Fig. 4(b-d)). We note that these values for $d$ are comparable to the linear dimension of the growth cone: $l$ = 2 to 5 μm. This range for $l$ was obtained from both fluorescence (Fig. 1) and AFM measurements [12] and it corresponds to the typical growth cone size reported in literature [1]. We will show that the axonal dynamics on these surfaces cannot be modeled by the simple OU process described by Eqs. (8-12) above. Inspired by previous work on directional cell migration [18-21] we introduce a neuronal growth model described by a non-linear Langevin equation:

$$\vec{a}(\vec{V},t) \equiv \frac{d\vec{V}}{dt} = \vec{a}_d(\vec{V},t) + \vec{\Gamma}(\vec{V},t) \tag{13}$$

where $\vec{a}_d(\vec{V},t)$ is the deterministic component of the axonal motion and the term $\vec{\Gamma}(\vec{V},t)$ represents the stochastic contributions.

We use the method of conditional averaging [18, 26] (see DATA ANALYSIS) to experimentally extract the deterministic and stochastic components in Eqn. (13) and their functional dependence on time, axonal speed and direction. Fig. 6 shows the variation of $a_{d,\|}(\vec{V},t)$ (the deterministic part of the acceleration parallel to the instantaneous direction of motion) with the growth cone speed $V$ for two different growth angles: $\theta = 0$ (Fig. 6(a)) and $\theta = \pi/2$ (Fig 6(b)). The continuous curves in these figures represent quadratic fit to the data points. We find that (for any angle $\theta$) the component of the deterministic acceleration parallel to the instantaneous direction of motion, $a_{d,\|}(\vec{V},t)$ is well approximated by a quadratic function of the growth cone *speed V*:

$$a_{d,\|}(V,\theta) = a_0 \cdot |\sin\theta| - \gamma_1 \cdot V - \gamma_2 \cdot V^2 \tag{14}$$

where $a_0$, $\gamma_1$, and $\gamma_2$ are three experimentally measurable parameters that characterize axonal dynamics on PDMS substrates with different geometries. These parameters are discussed below.

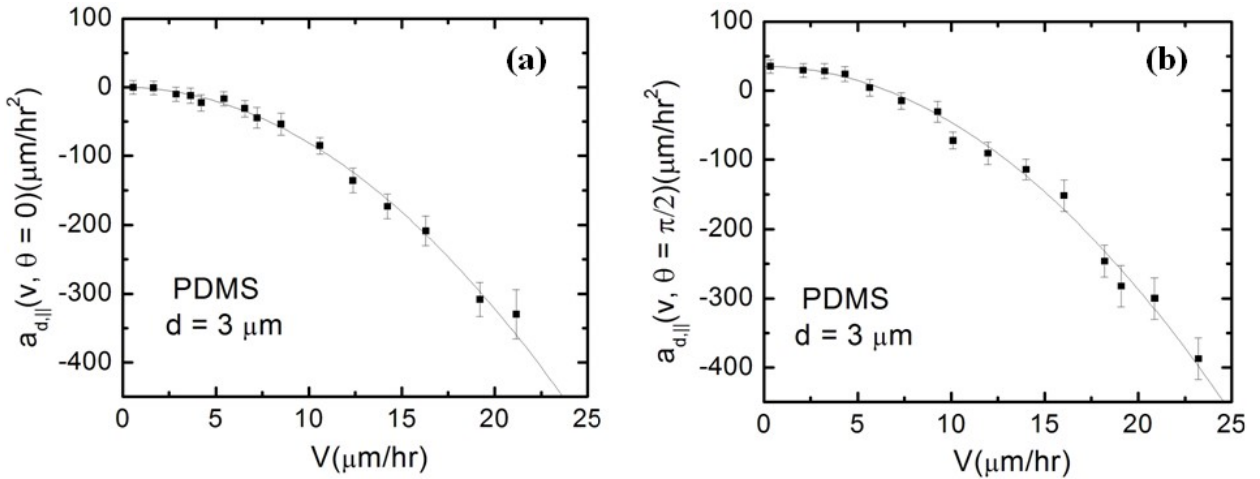

**FIG. 6.** Data points represent the variation of the parallel component of the deterministic acceleration with the growth cone speed for axons growing in $\theta = 0$ (a), respectively $\theta = \pi/2$ (b) directions (the growth angles are defined in Fig. 2(a)). The growth substrates are PDMS surfaces with $d$ = 3 μm. Each data point was obtained by conditional averaging the parallel component of acceleration according to Eq. (2), for a number N of growth cones between 10 and 40 per data point. Error bars indicate the standard error of the mean. The continuous curves represent fit to the data points with Eq. (14). The fit of the data points gives the following values for the growth parameters: $a_0 = (35.1 \pm 0.9)\,\mu m/hr^2$, $\gamma_1 = (0.09 \pm 0.06)\,hr^{-1}$, and $\gamma_2 = (0.8 \pm 0.1)\,\mu m^{-1}$.

The first term on the right hand side of the Eq. (14) shows that the angular dependence of the parallel component is described by the absolute value of the sine function (Fig. 7(a)). The absolute value $|\sin\theta|$ reflects the symmetry of the growth around the $x$ axis: the two distributions centered at $\theta = \pi/2$ and $\theta = 3\pi/2$ are symmetric with respect to the directions $\theta = \pi$ and $\theta = 0$, (as shown in Fig. 4(b-d)), which in turn means that there is no preferred direction along the pattern (i.e. the "up" and "down" directions in Fig. 1 and Fig. 2 are equivalent for neuronal growth). The fit of the data in Fig. (6) with Eq. (14) gives the following values for the three dynamical parameters that describe axonal growth on PDMS surfaces with $d = 3\,\mu m$ : $a_0 = (35.1 \pm 0.9)\,\mu m/hr^2$, $\gamma_1 = (0.09 \pm 0.06)\,hr^{-1}$, and $\gamma_2 = (0.8 \pm 0.1)\,\mu m^{-1}$. Remarkably, the experimental data shows that the parameter $\gamma_1$ (the coefficient of the linear term in $V$) has a constant value, which is independent of the neuronal growth surface and of the angle $\theta$ of the growth cone motion (Fig. S2(a) and Fig. S2(c) in the Supplemental Material [25]). Furthermore, the data shows that (within the experimental uncertainties): $\gamma_1 \approx \gamma_g$, i.e. the linear coefficient that describes neuronal growth on PDMS surfaces is equal to the constant damping coefficient for growth on glass. Thirdly, we find that for a given type of geometrical pattern (characterized by a constant spatial period $d$), the parameter $\gamma_2$ (the coefficient of the quadratic term in $V$) has a constant value that does not depend on the angle $\theta$ of the growth cone motion (Fig. S2(b) in the Supplemental Material [25]). The significance of these experimental findings will be presented in the DISCUSSION section.

The perpendicular component of the deterministic acceleration $a_{d,\perp}(\vec{V},t)$ was determined using Eqs. (3) and (4). The data shows that $a_{d,\perp}(\vec{V},t)$ is independent on the growth cone speed and depends only on the growth angle $\theta$ (Fig. 7(b) and Fig. S3 in the Supplemental Material [25]). The angular dependence is well approximated by the cosine function:

$$a_{d,\perp}(V,\theta) = a_1 \cdot \cos\theta \qquad (15)$$

where the parameter $a_1$ is a constant acceleration, for a given pattern spatial period $d$.

The magnitude $a_{d,\perp}(\theta)$ of the acceleration perpendicular to the instantaneous velocity has a maximum value when the direction of axonal growth is perpendicular to the surface pattern (i.e. for $\theta = 0$ and $\theta = \pi$), and it equals zero when the axon grows along the pattern ($\theta = \pi/2$ and $\theta = 3\pi/2$). This implies that the perpendicular component of acceleration $a_{d,\perp}(\vec{V},t)$ tends to *rotate the growth cone towards the direction* of the pattern. The parameter $a_1$ quantifies the magnitude of his effect. We note that this effect is similar to a previous result we have reported for neurons grown on directional nano-ppx surfaces [12]. In this previous work we have identified a deterministic torque that tends to align axons along certain preferred directions on the nano-ppx surface.

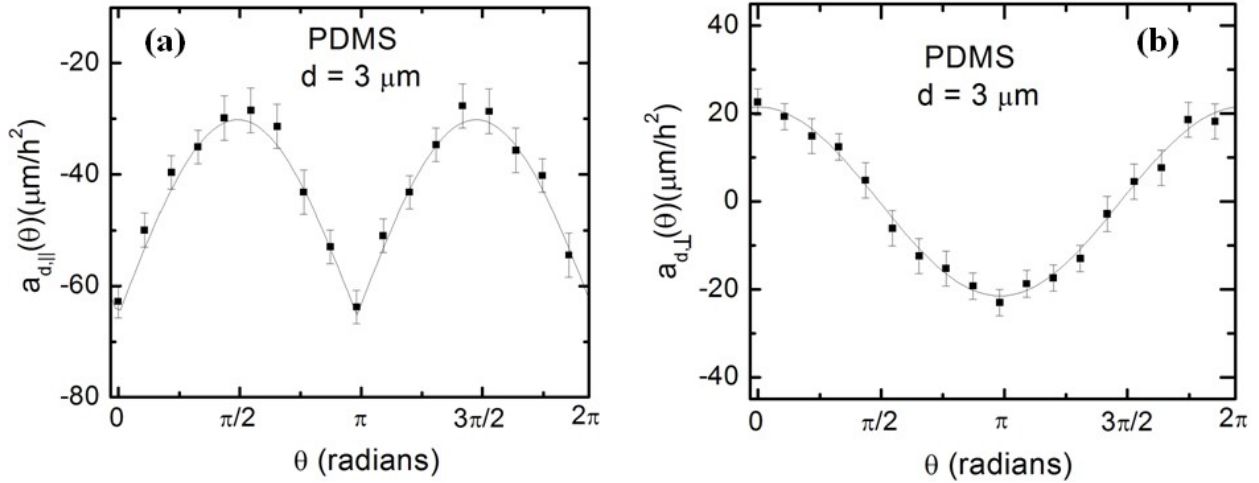

**FIG. 7.** Variation of the parallel (a), and perpendicular (b) components of the deterministic acceleration with the angle $\theta$ of axonal growth. The growth substrates are PDMS surfaces with $d = 3$ μm. Each data point was obtained by conditional averaging the parallel and perpendicular components of acceleration according to Eqs. (2-4), for a number N of growth cones between 30 and 110 per data point. Error bars indicate the standard error of the mean. The continuous curve in (a) is the plot of Eq. (14) with the parameters $a_0$, $\gamma_1$, and $\gamma_2$ found from the data shown in Fig. 6(a). The continuous curve in (b) represents the fit of the data points with Eq. (15). The data fit gives the following value for the angular parameter: $a_1 = (21.5 \pm 0.6)$ μm/hr$^2$.

From the fit to the data in Fig. 7(b) we find $a_1 = (21.5 \pm 0.6)$ μm/hr$^2$, for axonal growth on PDMS surfaces with $d = 3$ μm. By performing a similar analysis for neuronal growth on PDMS surfaces with different spatial periods: $d = 10$ μm, 5 μm, 1.5 μm and 0.5 μm we find the corresponding values for the deterministic parameters: $a_0$, $a_1$, $\gamma_1$, and $\gamma_2$ that control axonal dynamics on these surfaces. Fig. 8 shows the dependence of the parameters $a_0$, $a_1$ (Fig. 8(a)), and $\gamma_2$ (Fig. 8(b)) on the surface geometry quantified by the surface spatial periods $d$. The parameter $\gamma_1$ does not depend on $d$ (Fig. S2(c) in the Supplemental Material [25]).

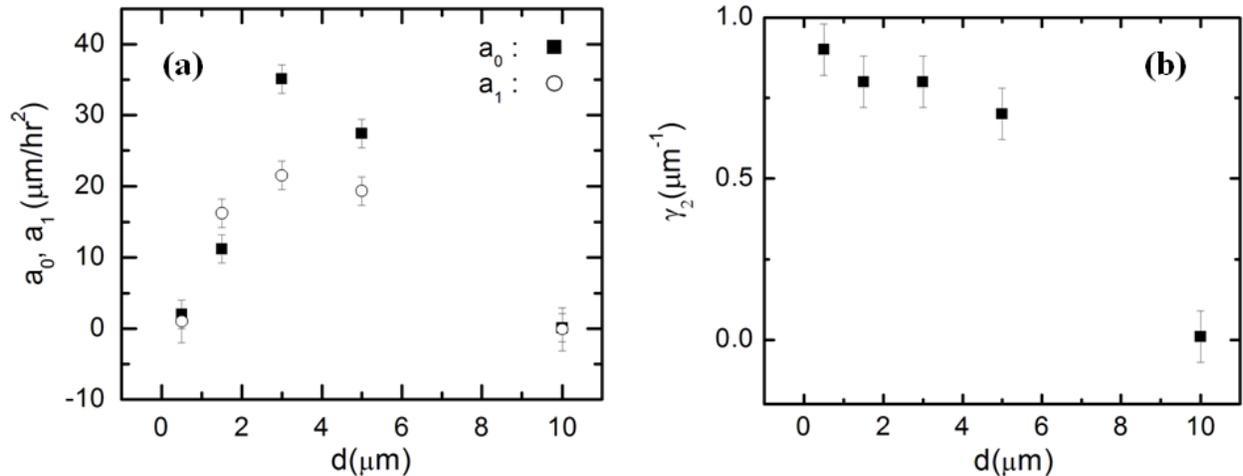

**FIG. 8.** (a) Variation of the of the acceleration parameters $a_0$ (squares) and $a_1$ (open circles) with the surface spatial period $d$. These parameters measure the magnitude of the angle

dependent terms of the deterministic acceleration: $a_0$ - for the parallel component, Eq. (14); $a_1$ for the perpendicular component, Eq. (15). (b) Variation of the coefficient of the quadratic term in speed with the surface spatial period. (a, b): Error bars indicate the standard error of the mean.

We conclude that the deterministic components of the acceleration for neuronal growth on PDMS surfaces where the pattern spatial periods $d$ matches the dimension $l$ of the growth cone: $d \approx l$, are described by four independent parameters. Three of these parameters are determined by the surface geometry. The angular dependence for each component is separated from the speed dependence. The component of the acceleration parallel to the instantaneous speed has a maximum value $a_0$ for growth cones moving along the surface pattern. Moreover, the parallel component has two additional damping terms: one linear and one quadratic in the growth cone speed, which lower the value of the acceleration along the surface pattern. The perpendicular component of the acceleration tends to align the axons along the pattern with an effective acceleration of constant magnitude $a_1$. All these terms contribute to the axonal alignment as we will show in the DISCUSSION section below.

To determine the stochastic terms in Eq. (13) we calculate the variance of the distributions for each component of the acceleration within each speed and angle interval (Eq. (5) and (6)) in DATA ANALYSIS). The data shows that the stochastic contributions for both parallel $\Gamma_\parallel$, and perpendicular $\Gamma_\perp$ components fluctuate around their average values with zero correlations (Fig. S4 in the Supplemental Material [25]). Consequently, we model the stochastic term in Eq. (13) as an uncorrelated Wiener process, with each component satisfying the conditions for Gaussian white noise with zero mean [19, 20, 27]:

$$\langle \Gamma_\parallel(t) \rangle = 0 \text{ and } \langle \Gamma_\parallel(t_1) \cdot \Gamma_\parallel(t_2) \rangle = \sigma_\parallel \cdot \delta(t_1 - t_2) \qquad (16)$$

$$\langle \Gamma_\perp(t) \rangle = 0 \text{ and } \langle \Gamma_\perp(t_1) \cdot \Gamma_\perp(t_2) \rangle = \sigma_\perp \cdot \delta(t_1 - t_2) \qquad (17)$$

where $\sigma_\parallel$, $\sigma_\perp$ quantify the strength of the noise, and $\delta(t_1 - t_2)$ is the Dirac delta – function.

Putting everything together (Eqs. (14-17)) we arrive at the conclusion that the conditional averaging of the experimental data imposes the following two equations of motion for the growth cones on patterned PDMS surfaces with $d$ comparable to the growth cone dimension:

$$a_\parallel(V, \theta, t) \equiv \left(\frac{d\vec{V}}{dt}\right)_\parallel = a_0 \cdot |\sin\theta| - \gamma_1 \cdot V - \gamma_2 \cdot V^2 + \Gamma_\parallel(t) \qquad (18)$$

$$a_\perp(V, \theta, t) \equiv \left(\frac{d\vec{V}}{dt}\right)_\perp = a_1 \cdot \cos\theta + \Gamma_\perp(t) \qquad (19)$$

**Neuronal growth on micro-patterned PDMS surfaces with spatial periods $d = 0.5$ μm, and $d = 10$ μm.** Experimental data shows that there is no directional alignment for neurons grown on PDL coated PDMS surfaces with the pattern spatial periods $d = 0.5$ μm, and $d = 10$ μm (Fig. S1 in the Supplemental Material [25]). For these types of surfaces the values of the measured parameters that describe the angular dependence of the acceleration (Eqs. (18) and (19)) are close to zero: $a_0 \approx 0$, and $a_1 \approx 0$ (Fig. 8(a)). Furthermore, the values for the coefficients of the quadratic term in $V$ in Eq. (18) for these surfaces are: $\gamma_2 = (0.9 \pm 0.08)\,\mu\text{m}^{-1}$ for $d = 0.5$ μm, respectively $\gamma_2 = (0.01 \pm 0.08)\,\mu\text{m}^{-1}$ for $d = 10$ μm (Fig 8(b)). The coefficients of the linear terms in $V$ in Eq. (18) are equal (within the experimental uncertainties) to the constant damping coefficient for growth on glass $\gamma_1 \approx \gamma_g$ (Fig. S2(c) in the Supplemental Material [25]).

The values of the parameters measured for neuronal growth on these PDMS surfaces imply that the growth cone dynamics on PDMS surfaces with the pattern spatial periods $d \gg l$ and $d \ll l$ is similar to the dynamics of growth cones on glass, and thus *it is described* by an OU process. This approximation is exact for PDMS surfaces with $d = 10$ μm for which Eqs. (18) and (19) are simply reduced to a linear Langevin equation (Eq. (8)) with the *same value* for the constant damping coefficient $\gamma_g$ (the quadratic term in $V$ is zero). For neuronal growth on PDMS surfaces with $d = 0.5$ μm the coefficient $\gamma_2$ of the quadratic term in $V$ is nonzero, and the Langevin Eq. (18) is nonlinear. However, even in this case the nonlinear effects are small, there is no angular dependence of the acceleration ($a_0 \approx 0$), no alignment term ($a_1 \approx 0$), and no terminal velocity (see below). Thus the OU process provides a very good approximation for describing neuronal growth dynamics on both types of PDMS surfaces ($d = 0.5$ μm, $d = 10$ μm), which are incommensurate with the growth cone dimensions.

## VI. DISCUSSION

The experimental data for neurons grown on PDL coated glass shows that the axonal dynamics on these surfaces is governed by an Ornstein-Uhlenbeck process, i.e. a linear Langevin equation with Gaussian white noise (Eq. (8)). The OU process with Gaussian white noise, which is inspired by the study of the Brownian motion, represents the simplest stochastic model used for describing cellular motility. It has been successfully used for modeling the dynamics of many types of cells including endothelial cells [22], human granulocytes [19], fibroblasts and human keratinocytes [20], as well as cortical neurons [12, 16, 17]. The values we have obtained for the diffusion coefficient and the characteristic speed for growth cones on glass are comparable with the corresponding values reported for human peritoneal mesothelial cells [28], one order of magnitude smaller than the values reported for human keratinocytes [20], and for endothelial cells [22], and about two orders of magnitude smaller than the corresponding values reported for glioma cells [29]. These results are consistent with the relatively slower dynamics expected for growth cones as they move to form connections and to wire up the nervous system [1, 2].

As we have noted above the linear dimensions for the growth cones of the cortical neurons are in the range $l = 2$ to $5$ μm. The experimental data (Fig. 1(b-c), and Fig. 4(b-c)), shows that axons display maximum alignment along the PDMS patterns for substrates with $d = 3$ μm and $d = 5$ μm, i.e. for surfaces where *the pattern spatial period matches the linear dimension of the growth cone*: $d \approx l$. The PDMS surfaces with $d = 1.5$ μm (for which $d < l$) show a lower degree of axonal alignment (Fig. 1(d), and Fig. 4(d)). A detailed analysis of the data (previous section) shows that, unlike the growth on glass, the dynamics of growth cones on these PDMS surfaces cannot be described by an OU process. We found that Eqs. (18) and (19) summarize the growth dynamics on PDMS surfaces with $d \approx l$ (instead of Eq. (8) for the OU process on glass).

Equations (18) and (19) have several remarkable features. First, they show that the acceleration parallel to the instantaneous direction of motion is decoupled from the acceleration perpendicular to this direction (including the stochastic terms). Second, the angular dependence of the motion for each direction can be separated from the dependence on speed. The deterministic part of the motion in the parallel direction (Eq. (18)) has a sine dependence on the angle multiplied by a constant magnitude factor $a_0$, as well as linear and quadratic dependence on speed. Mathematically this is similar to the motion of an object in gravitational field with both linear and quadratic air resistance, as we will discuss below. Thirdly, the deterministic term in the equation for the perpendicular direction has a cosine dependence on the angle. This term has a maximum value $a_1$ for the direction perpendicular to the direction of the geometrical pattern ($\theta = 0$ and $\theta = \pi$) and equals zero for directions along the pattern ($\theta = \pi/2$ and $\theta = 3\pi/2$). Therefore the growth cone tends to rotate as it extends and aligns with the pattern. It is the combination of this effect with the fact that the growth cone reaches a terminal velocity along the direction pattern (see below) that is ultimately responsible for the high degree of alignment between the axons and the geometrical patterns.

We note that for a given growth angle $\theta$, the deterministic part of the acceleration parallel to the instantaneous velocity (Eq. 19) has the same mathematical form as the equation describing the vertical motion of an object in gravitational field, subject to both linear and quadratic air resistance [30]. For growth angles $\theta \neq 0$, Eq. (18) implies the existence of a deterministic terminal speed (the contribution from stochastic terms averages to zero) given by the condition: $a_{d,\parallel}(V,\theta,t) = \left\langle \left| \frac{d\vec{V}}{dt} \right|_\parallel \right\rangle = 0$. The terminal speed depends on the pattern spatial period $d$, and (for a given $d$) it has a maximum value for axonal growth along the pattern direction: $\theta = \pi/2$ and $\theta = 3\pi/2$ (Fig. S5 in the Supplemental Material [25]). For example, for growth on PDMS surfaces with $d$ = 3 μm, we get: $V_{ter} = (6.5 \pm 0.6)$ μm/hr (see Supplemental Material [25]). This value is a factor ~ 3 larger than the characteristic speed for the random OU growth on glass (Eq. (12)), consistent with longer axonal lengths measured on PDMS surfaces.

Fig. S1 (Supplemental Material [25]) shows that, in contrast with the case discussed above, there is no alignment for neuronal growth on PDMS surfaces with $d$ = 0.5 μm and $d$ = 10 μm. For these surfaces the pattern spatial period $d$ is either *much smaller or much larger* than the linear dimension $l$ of the growth cone. There is no deterministic term that tends to align axonal growth along the direction of the patterns, and the terminal speed for axons growing along the pattern is $V_{term} \approx 0$ (Fig. S5 in the Supplemental Material [25]). The growth cone dynamics on these surfaces is described by an OU process with the same parameters as in the case of neuronal growth on glass (Eq. (8)).

Eqs. (18) and (19) (as well as Eq. (8) for the growth on glass) contain terms described by Gaussian white noise that characterizes stochastic changes in speed. The Gaussian white noise is a general characteristic of cellular motion, and it reflects the stochastic nature of both the extra cellular (neuron-neuron) signaling [1, 2, 7, 8], as well as the intra-cellular processes, such as: the stochasticity of biochemical reactions taking place in the growth cone, polymerization rates of microtubules and actin filaments, and the formation of lamellipodia and filopodia [1, 4, 16]. Theoretical models using uncorrelated Gaussian white noise have been used to describe the dynamics of endothelial cells [22], human granulocytes [19], fibroblasts and human keratinocytes [20], as well as cortical neurons [12, 16, 17]. We mention that the description of the dynamics for some other types of eukaryotic cells, e.g. *D discoideum*, requires the introduction of speed-dependent multiplicative noise, which suggests a nonlinear character of the chemotactic mechanisms in these cells [18]. In the case of neurons however, we conclude that the stochastic part of neuronal dynamics is described by Gaussian white noise, and it is not influenced by geometry, velocity or angle of motion.

**Role of geometry on neuronal growth on micro-patterned PDMS surfaces.** The axonal alignment on PDMS surfaces with $d \approx l$ is completely determined by four measurable parameters $a_0$, $a_1$, $\gamma_1$, and $\gamma_2$, which characterize the deterministic components of the growth cone acceleration. Three of these parameters depend on the surface geometry (quantified by the pattern spatial period $d$), while the parameter $\gamma_1$ is surface independent. We emphasize that the directional motion of the growth cone results from *two* combined effects: a) growth cones are more likely to move in the directions parallel with the surface patterns (cosine dependence in Eq. (19) tends to rotate the growth cone along these directions); and b) growth cones that are moving along the direction of the surface patterns have maximum speed (Eq. (18) and Fig. S5 in the Supplemental Material [25]). These results are consistent with contact-guidance behavior that we and other groups have previously reported for neurons grown on different types of patterned surfaces [12, 31-34]. Contact guidance is defined as the ability of some cells to orient their motion in response to surface geometrical structures. It has been observed for many types of cells including granulocytes, fibroblasts, and tumor cells [31-35]. Growth cones have several different types of surface receptors and membrane curvature sensing proteins involved in surface adhesion, and locomotion including amphipathic helices and bin-amphiphysin-rvs (BAR) - domain containing proteins [1, 5, 31]. In the case of contact-guidance locomotion it has been shown that the degree of directional cell motility increases proportionally with the density of anchored surface receptors [31-35]. An important parameter for contact guidance is the ratio between the cell size and the characteristic dimensions of the surface geometrical features [31]. This parameter determines the surface density of surface receptors, which mediate adhesion and mechanotransduction between the cell cytoskeleton and the substrate. We hypothesize that for neurons grown on PDMS surfaces with $d \approx l$ (i.e. where the linear dimension of the growth cone matches the pattern spatial period) the growth cone "wraps tightly" around the surface features, which results in a minimum contact area and thus maximum density of surface receptors. Previous reports have shown that the maturation of the surface receptor and focal adhesion points respond to external forces, including cell-substrate traction forces [36]. Thus the focal contacts on a filopodium wrapped over a geometrical feature (ridge) with high curvature (see AFM images of the PDMS patterns in Fig. 2) will undergo higher forces than those contacting a lower curvature feature. Furthermore, microtubules and actin filaments inside the growth cone act as stiff load-bearing structures that provide resistive forces to the bending of the filopodia. Together these effects will ultimately lead to axonal alignment along the PDMS surface pattern when the pattern spatial period is comparable to the size of the growth cone.

We conclude that the theoretical model given by Eqs. (18) and (19) represents the simplest non-linear generalization of the OU process, that: 1) fully accounts for the experimental data of neuronal growth on PDMS surfaces; 2) has a minimum number of phenomenological parameters that account for the cell-surface interactions; and 3) allows for meaningful comparisons with the simpler case of linear Langevin dynamics that describes neuronal dynamics on glass, as well as for comparison with the dynamics of other types of cells reported in literature. The model predicts characteristic speeds for neuronal growth on surfaces with uniform geometries (e.g. glass and PDMS with $d \gg l$ and $d \ll l$), terminal velocities for neuronal growth on surfaces with $d \approx l$, deterministic torque that tends to align axons along certain preferred directions along the surface, and cross over between linear (OU) and nonlinear dynamics, which depend on surface geometry. We hypothesize that these could be general features of cellular motility in various environments with inhomogeneous physical and chemical properties. Evidence for this hypothesis comes from previous studies of neuronal growth on surfaces with various geometries, textures and biochemical properties [3, 7, 9-17], as well as from motility studies for other types of cells [18-22]. In addition, the model could be further extended to account for the explicit dependence of the phenomenological parameters on the surface geometrical properties (such as pattern period *d* for the PDMS surfaces presented here). This will require measuring cell-surface coupling forces (using for e.g. traction force microscopy) and quantifying the density of cell surface receptors (using fluorescence techniques) that determine axonal contact guidance dynamics. In principle these future studies will allow to quantify the influence of environmental cues (geometrical, mechanical, biochemical) on neuronal growth, and to correlate the observed growth dynamics with cellular processes (cytoskeleton dynamics, cell-surface interactions, cell-cell communication etc.).

## VII. CONCLUSIONS

In this paper, we have used stochastic analysis to model neuronal growth on PDL coated glass and micro-patterned PDMS substrates coated with PDL. We have shown that the experimental data for neurons grown on glass, and on PDMS substrates with pattern spatial periods which are large, or small, compared to the dimension of the growth cone are well-described by Ornstein-Uhlenbeck (OU) processes (linear Langevin equations with white noise). On the other hand, neuronal growth on PDMS surfaces where the pattern spatial period matches the dimension of the growth cone ($d \approx l$) cannot be described by an OU process. To our knowledge this is the first description of the cross over between linear (OU) and nonlinear behavior for the same type of cell, which is controlled only by the surface geometry.

These results are consistent with contact-guidance phenomenon for neuronal growth. We conclude that the growth behavior of axons on these surfaces is determined solely by *one external parameter* (pattern period) which characterizes the surface geometry. The model is general and could be applied to neurons cultured on other types of substrates with different geometrical features as well as to neuronal growth *in vivo*. Moreover it could be applied to the motion of other types of cells in controlled environments including electric fields, surfaces with different stiffness, or biomolecular cues with different concentration gradients.

**Acknowledgement:** The authors thank Dr. Steve Moss's laboratory at Tufts Center of Neuroscience for providing embryonic rat brain tissues. The authors gratefully acknowledge financial support for this work from Tufts Summer Scholars (IY), and Tufts Faculty Award (FRAC) (JMVB, CS).

# Supplemental Material for

## *Role of geometrical cues in neuronal growth*

Joao Marcos Vensi Basso[1], Ilya Yurchenko[1], Marc Simon[1], Daniel J. Rizzo[1,2], Cristian Staii[1,*]

1. Department of Physics and Astronomy, Center for Nanoscopic Physics, Tufts University, Medford, Massachusetts 02155, *USA*

2. Current address: Department of Physics, University of California, Berkeley, California, 94720, *USA*

[*] Corresponding Author: Prof. C. Staii, E-mail: Cristian.Staii@tufts.edu

In this supplemental material we provide additional experimental details about surface preparation and cell culture (S1), as well as fluorescence and atomic force microscope imaging (S2). We then present fluorescence images and angular distributions for neuronal growth on PDMS surfaces with the pattern spatial periods $d = 0.5$ μm, and $d = 10$ μm (S3, Fig. S1). We also present additional experimental data (S3) for: 1) the variation of the measured linear and quadratic speed coefficients $\gamma_1$, and $\gamma_2$ (see Eqs. (18) and (19) in the main text) with the growth angle and the pattern spatial period (Fig. S2); 2) the variation of the acceleration parameter $a_1$ with the growth cone speed (Fig. S3); and 3) the variation of stochastic terms $\Gamma_\parallel$, and $\Gamma_\perp$ with the growth angle and speed (Fig. S4). We also show that the cross-correlation terms are much smaller than the stochastic terms (S3), as claimed in the DATA ANALYSIS section (main text). Finally, we present a brief discussion of the terminal velocity for the growth cones moving along the surface patterns (S4) and show the variation of terminal velocity with the pattern spatial period (Fig. S5).

## S1. Surface preparation and cell culture

To fabricate micro-patterned substrates we start with 20mL polydimethylsiloxane (PDMS) solution (Silgard, Dow Corning) and pour it over diffraction gratings with slit separations of: 10 μm, 5 μm, 3 μm, 1.5 μm and 0.5 μm and total surface area 25 x 25 mm$^2$ (Scientrific Pty. and Newport Corp. Irvine, CA). The PDMS films were left to polymerize for 48 hrs at room temperature, then peeled away from the diffraction gratings and cured at $55^0$ C for 3 hrs. We use AFM imaging to ensure that the pattern was successfully transferred from the diffraction grating to the PDMS surface (Fig. 2). The result is a series of periodic patterns (parallel lines with crests and troughs) with constant distance $d$ between two adjacent lines (Fig. 2). The surfaces were then glued to glass slides using silicone glue, and dried for 48 hours. Next, each surface was cleaned with sterile water and spin-coated with 3 mL of Poly-D-lysine (PDL) (Sigma-Aldrich, St. Louis, MO) solution of concentration 0.1 mg/mL. The spinning was performed for 10 minutes at 1000 RPM. Prior to cell culture the surfaces have been sterilized using ultraviolet light for 30 minutes.

Cortical neurons have been obtained from rat embryos (day 18 embryos obtained from Tufts Medical School). The brain tissue protocol was approved by Tufts University Institutional Animal Care Use Committee and complies with the NIH guide for the Care and Use of Laboratory Animals. The cortices have been incubated in 5 mL of trypsin at 37ºC for 20 minutes. To inhibit the trypsin we have used 10 mL of soybean trypsin inhibitor (Life Technologies). Next, the neuronal cells have been mechanically dissociated, centrifuged, and the supernatant was removed. After this step the neurons have been re-suspended in 20 mL of neurobasal medium (Life Technologies) enhanced with GlutaMAX, b27 (Life Technologies), and pen/strep. Finally, the neurons have been re-dispersed with a pipette, counted, and plated on PDL coated glass, or PDL coated PDMS substrates, at a density of 5,000 cells/cm$^2$.

## S2. Fluorescence and AFM Imaging

For fluorescence imaging the cortical neurons cultured on glass or PDMS surfaces, were rinsed with phosphate buffered saline (PBS) and then incubated for 30 minutes at 37ºC with 50 nM Tubulin Tracker Green (Oregon Green 488 Taxol, bis-Acetate, Life Technologies, Grand Island, NY) in PBS. The samples were then rinsed twice with PBS and re-immersed in PBS solution for imaging. Fluorescence images were acquired using a standard Fluorescein isothiocyanate -FITC filter: excitation of 495 nm and emission 521 nm. Axon outgrowth was tracked using ImageJ (National Institute of Health). To obtain the angular distributions (Fig. 4 and Fig. S1) all axons have been tracked and then partitioned into segments of 20 μm in length. We have then recorded the angle that each segment makes with the $x$ axis (Fig. 2), and the results were plotted as angular histograms (Fig. 4 and S1). All surfaces were imaged using an MFP3D Atomic Force Microscope (AFM), equipped with a BioHeater closed fluid cell, and an inverted Nikon Eclipse Ti optical microscope (Micro Video Instruments, Avon, MA). The AFM topographical images of the surfaces were obtained using the AC mode of operation, and AC 160TS cantilevers (Asylum Research, Santa Barbara, CA). Surfaces were imaged both before and after neuronal culture, and no significant change in topography was observed.

## S3. Additional Experimental Data (see the following pages)

## S3. Additional Experimental Data

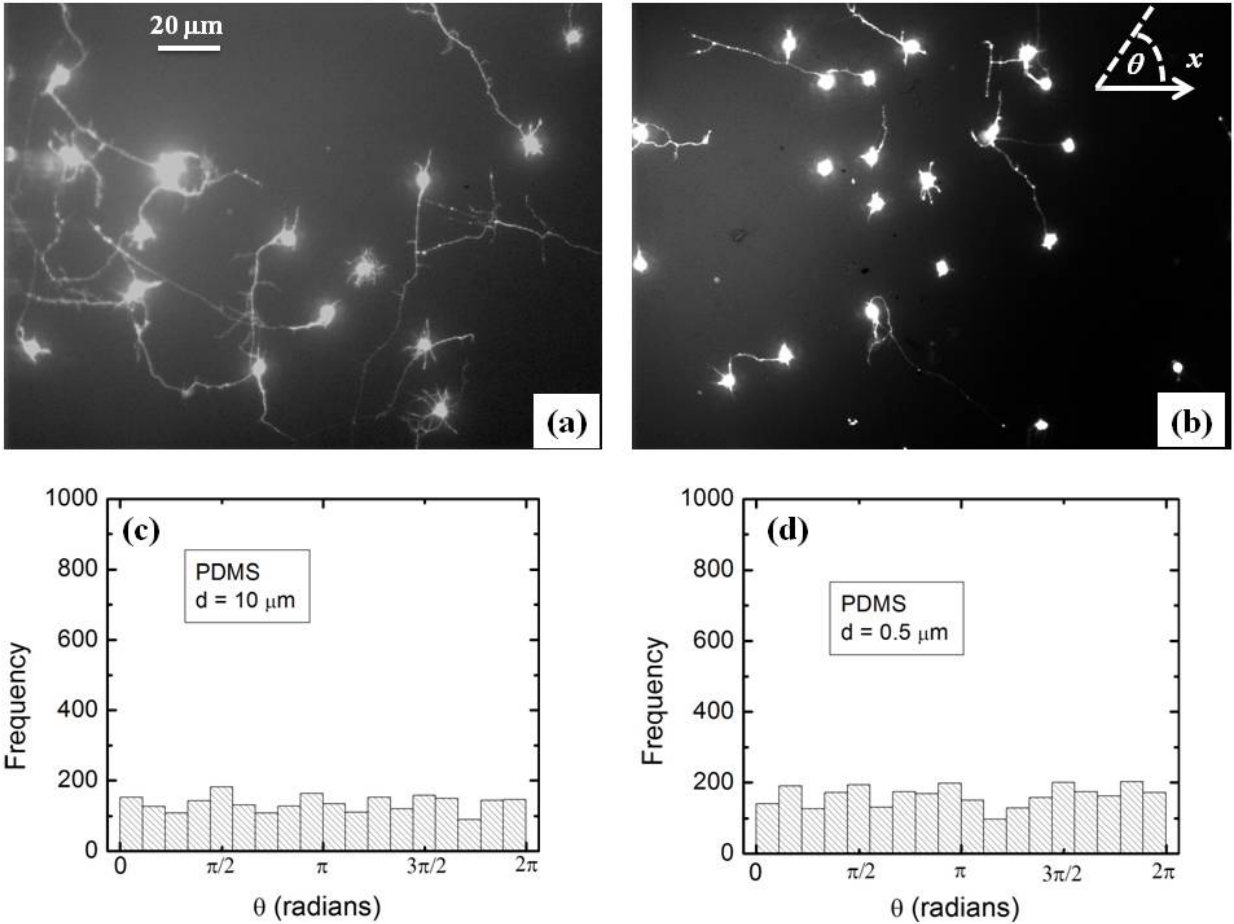

**FIG. S1**. (a-b): Examples of cultured cortical neurons on PDL coated PDMS surfaces with pattern spatial period $d = 10$ μm in (a), and $d = 0.5$ μm in (b). (c-d): Experimental angular distributions for axonal growth on PDMS with $d = 10$ μm in (c), and $d = 0.5$ μm in (d). The vertical axis (labeled Frequency) represents the number of axon segments, each one of 20 μm in length (see DATA ANALYSIS section). All angles are measured with respect to the *x* axis, defined as the axis perpendicular to the direction of the PDMS patterns (Fig. 2). The total number of axon segments is N= 2590 in (a) and N= 2640 in (b).

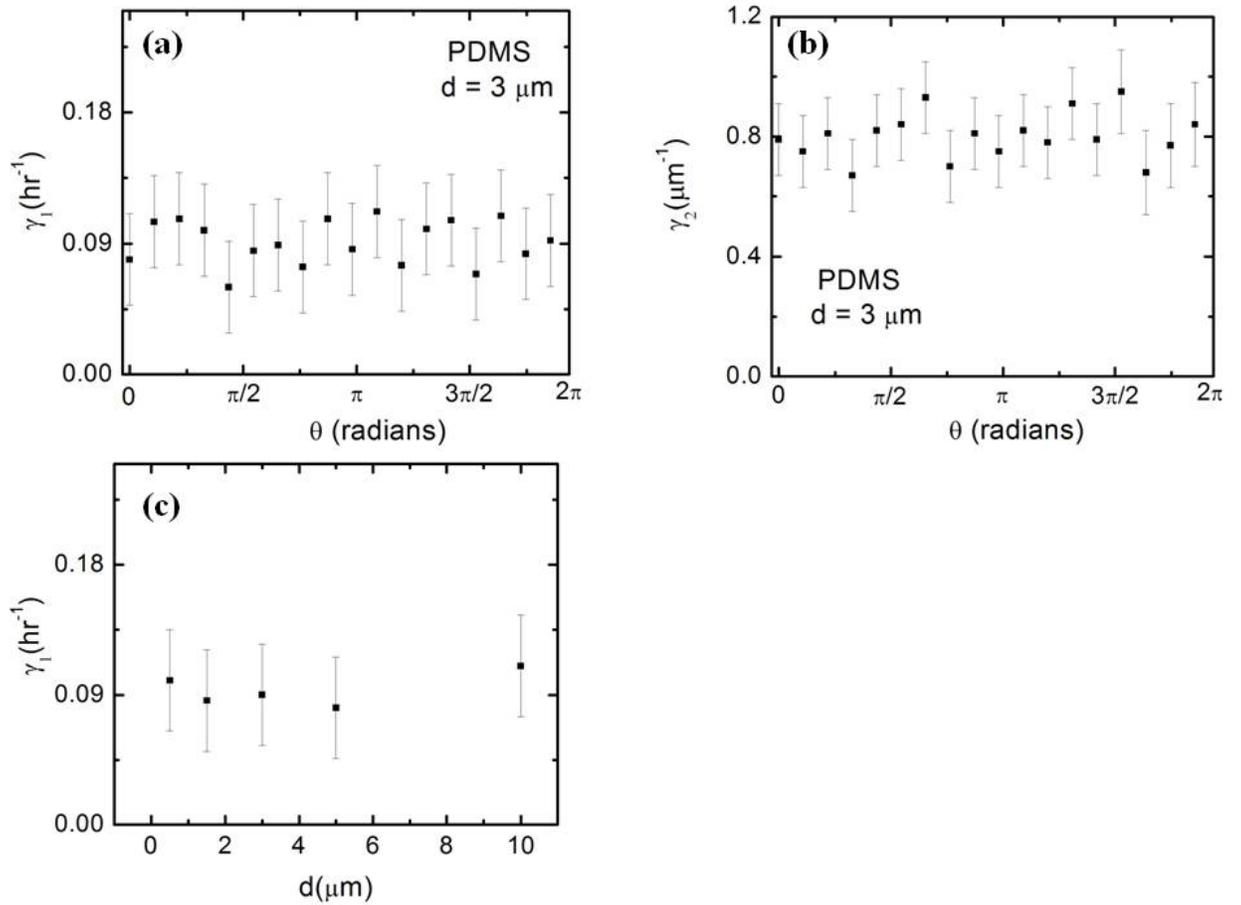

**FIG. S2.** Variation of the linear (a) and quadratic (b) speed coefficients with the direction of axonal growth. The growth substrates are PDMS surfaces with pattern spatial period $d = 3$ μm. Both coefficients are independent of the growth angle. (c) Variation of the coefficient of the linear term in speed with the surface spatial period. This coefficient is independent of the pattern spatial period. (a - c): Error bars indicate the standard error of the mean.

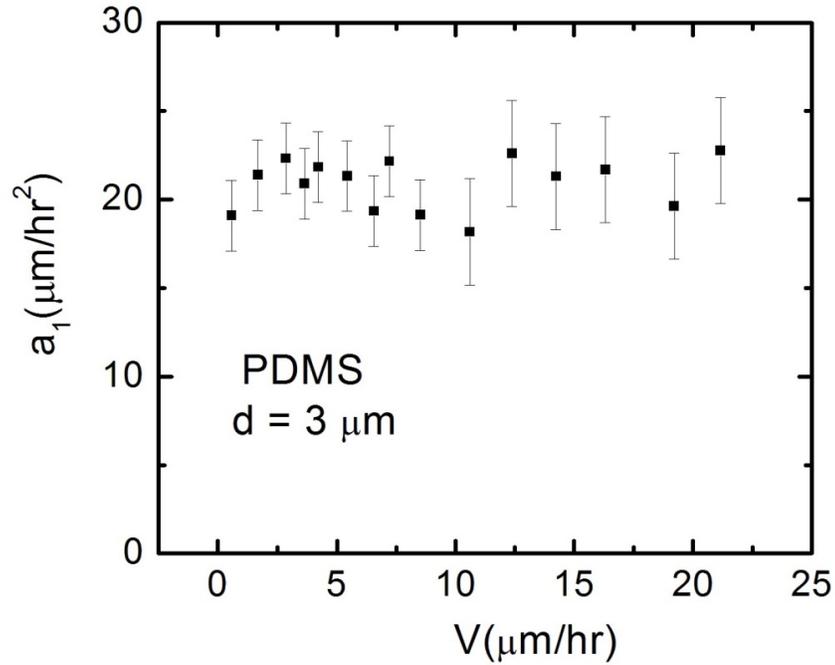

**FIG. S3.** Acceleration parameter $a_1$ as a function of growth cone speed, for neurons grown on PDMS surfaces with pattern spatial period $d = 3$ μm. The parameter $a_1$ is the maximum value of the perpendicular component of the deterministic acceleration, obtained for growth cones moving perpendicular with respect to the surface pattern ($\theta = 0$ and $\theta = \pi$). The data shows that the perpendicular component of the deterministic acceleration is independent of the growth cone speed. Error bars indicate the standard error of the mean.

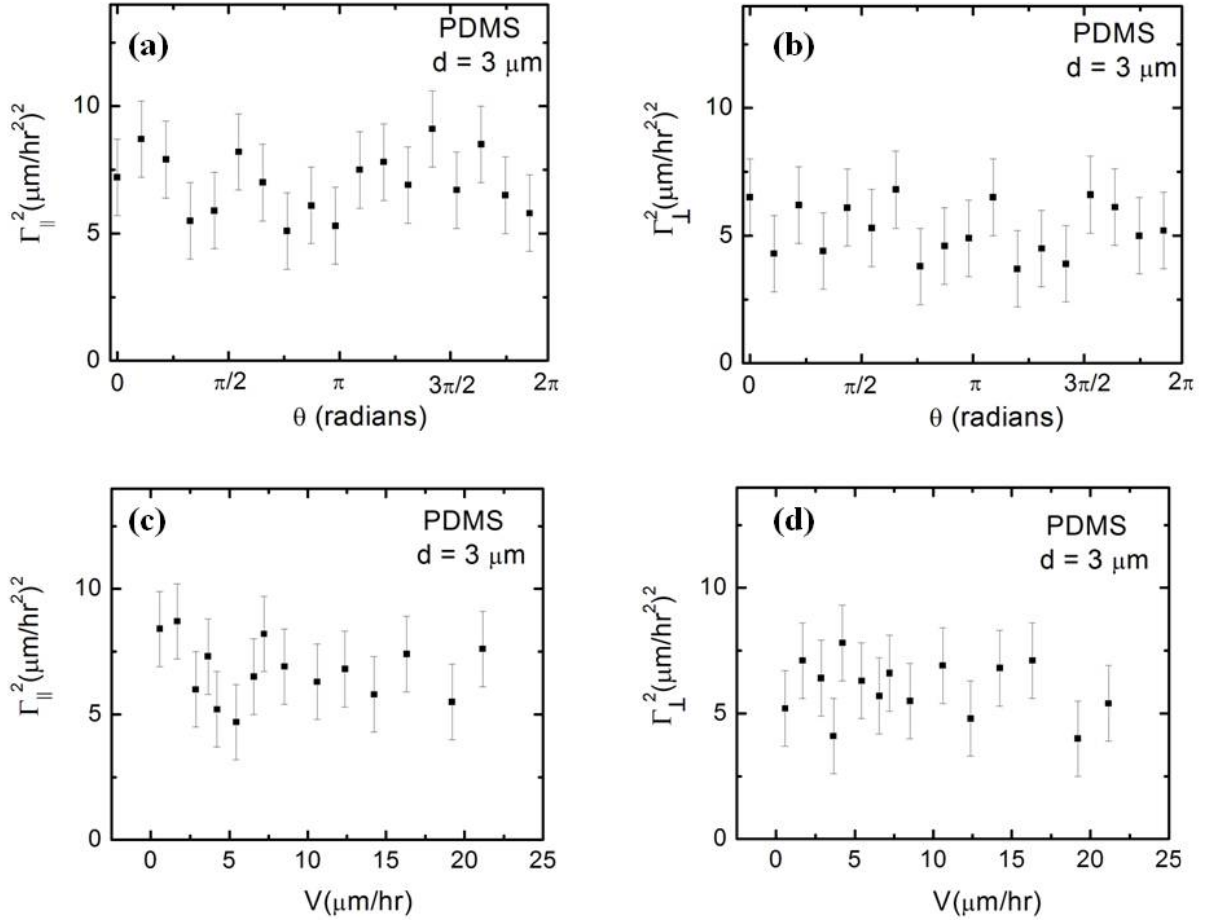

**FIG. S4**. (a-b): Stochastic components of the parallel (a), and perpendicular (b) accelerations as a function of axonal growth angle. (c-d) Stochastic components of the parallel (c), and perpendicular (d) accelerations as a function of the growth cone speed. Error bars represent a 95% confidence interval of a binomial distribution. The stochastic components are independent of the growth cone speed and growth angle.

**Cross-correlation terms**

The cross correlation terms are calculated according to [18]:

$$\Gamma_{\parallel}\Gamma_{\perp} = \left\langle \left(a_{\parallel}(\vec{V},t) - \langle a_{\parallel}(\vec{V},t)\rangle\right) \cdot \left(a_{\perp}(\vec{V},t) - \langle a_{\perp}(\vec{V},t)\rangle\right) \right\rangle_C \quad (S1)$$

The values we obtain for these cross-correlation terms are in the range 0.01-0.2 $(\mu m/hr^2)^2$, i.e. one - two order of magnitude smaller than the stochastic terms shown in Fig. S4. We therefore neglect the cross-correlation terms in the components of the acceleration, as stated in the main text.

## S4. Terminal velocity for growth cones moving along the surface patterns

The terminal velocity of the growth cone is found from the condition:

$$a_{d,\parallel}(V,\theta,t) = \left\langle \left|\frac{d\vec{V}}{dt}\right|_{\parallel} \right\rangle = 0$$

Imposing this condition to Eq. (18) we get the following expression for the terminal velocity:

$$V_{ter} = \sqrt{\frac{a_0}{\gamma_2}\cdot|\sin\theta| + \frac{\gamma_1^2}{4\gamma_2^2}} - \frac{\gamma_1}{2\gamma_2} \qquad (S2)$$

Eqn. (S2) shows that the growth cones reach terminal velocity only for growth angles $\theta \neq 0$. For a given $d$ the terminal velocity has a maximum value for axonal growth along the pattern direction: $\theta = \pi/2$ and $\theta = 3\pi/2$. The terminal speed depends on the pattern spatial period $d$ through the parameters $a_0$, and $\gamma_2$. Fig. S5 shows the variation of the terminal velocity (obtained for axons growing in the $\theta = \pi/2$ direction) with the pattern spatial period $d$.

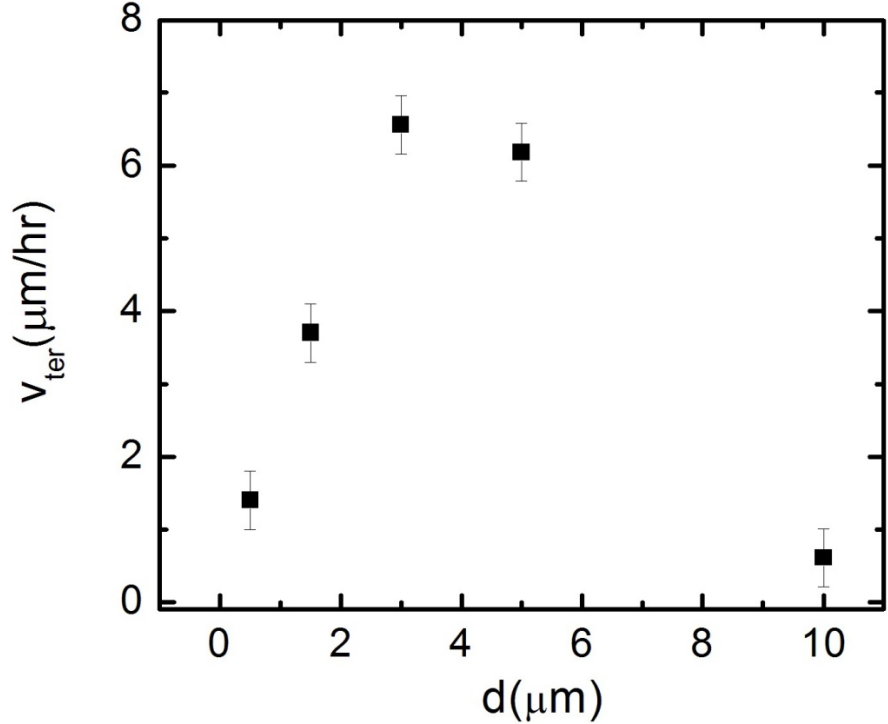

**FIG S5.** Variation of the terminal velocity of the growth cone (obtained for $\theta = \pi/2$) with the pattern spatial period $d$. The terminal velocity has maximum values for PDMS surfaces with $d = 3$ μm and $d = 5$ μm, i.e. for surfaces where axons display maximum alignment.